\documentclass[10pt,runningheads]{llncs}
\begin{document}
\titlerunning{An ID-based Ring Signature Scheme ...}
\title{ID-based Ring Signature and Proxy Ring Signature Schemes from Bilinear Pairings}
\author{Amit K Awasthi\inst{1} \and Sunder Lal\inst{2}}
\institute{Department of Applied Science \\ Hindustan College of Science and Technology \\Farah, Mathura -281122, INDIA \\ \email{awasthi\_hcst@yahoo.com} \and Department of Mathematics \\ I. B. S. Khandari, Agra - INDIA } 
\maketitle
\begin{abstract} In 2001, Rivest et al. firstly introduced the concept of ring signatures. A ring signature is a simplified group signature without any manager. It protects the anonymity of a signer. The first scheme proposed by Rivest et al. was based on RSA cryptosystem and certificate based public key setting. The first ring signature scheme based on DLP was proposed by Abe, Ohkubo, and Suzuki. Their scheme is also based on the general certificate-based public key setting too. In 2002, Zhang and Kim proposed a new ID-based ring signature scheme using pairings. Later Lin and Wu proposed a more efficient ID-based ring signature scheme. Both these schemes have some inconsistency in computational aspect. 

In this paper we propose a new ID-based ring signature scheme and a proxy ring signature scheme. Both the schemes are more efficient than existing one. These schemes also take care of the inconsistencies in above two schemes.

\end{abstract}
\section{Introduction}
The concept of ring signature was introduced by Rivest, Shamir and Tauman in \cite{Rivest01:HLS}. The ring signature allows a user from a set of possible signers to convince the verifier that the author of the signature belongs to the set but identity of the author is not disclosed. The ring signature may be considered to be a simplified group signature which consists of only users without the managers. It protects the anonymity of a signer since the verifier knows only that the signature comes from a member of a ring, but doesn't know exactly who the signer is. There is no way to revoke the anonymity of the signer. 
\paragraph{} Unlike the group signature schemes 
 the ring signature scheme requires neither a group manager, nor a setup procedure, nor the action of non-signing members. For signing any message $m$, the signer may choose random set of other possible signers including himself, to produce a valid ring signature. This signature does not reveal the identity of the signer but it may be verified with this signature that the signer belong to the possible signers set. There is no revocation manager. This allows unconditional anonymity of signer.
\paragraph{} \emph{Ring Signature with Proxy Signatures:} The proxy signature scheme was introduced by Mambo et al. in 1996. These allow a proxy signer to sign on behalf of an original signer. After Mambo et al. many proxy signature schemes have been proposed \cite{KPW97,LCK03,Lee00:EPr-ProtectedProxySignature}. Proxy signature may be combined with other special signatures to obtain new type of the proxy signatures. Various schemes like multi-proxy signature scheme \cite{HSh01:ASMPSS}, threshold proxy signature scheme, proxy blind signature scheme \cite{awa} etc. have been proposed.
\paragraph{} Suppose an original signer delegates its signing capability to a number of proxy signers, such that any proxy signer may produce a valid proxy signature for some message $m$. To achieve anonymity of these proxy signers we may use ring signatures. We, therefore, combine the idea of proxy signature with ring signature and get a new type of signature -\textit{ proxy ring signature}.
\paragraph{} In 1984, Shamir \cite{Sha84} introduced ID-based encryption and signature schemes to simplify key management procedures in certificate-based public key setting. Recently, many bilinear pairings based ID-based signature schemes were developed. \cite{Bon01,BLS01,Boy03,ChC03,CC01,Hes03,Hes02,ZK02,ZNL03}
\paragraph{} In this paper we propose a new ID-based ring signature scheme and also an ID-based proxy ring signature scheme. Both schemes are more efficient than existing schemes.
\section{Overview of Ring Signatures}
In this section we follow formalization proposed by Rivest et al. \cite{Rivest01:HLS}
\begin{definition}
Assume that each user has a secret key $S_i$ and its corresponding public key $P_i$. Let $<P_r>$ denotes the set of possible signer where $r$ is number of users listed in the set. Then ring signature scheme consists of the following algorithms --
\begin{itemize} 
\item{Ring Sign-} A probabilistic algorithm which takes a message $m$, secret key $S_k$ of signer, and the possible signers set $<P_r>$ as input and produces a ring signature $\sigma$ for the message $m$
\item{Ring Verify-} A deterministic algorithm which takes a message $m$, the possible signers set $<P_r>$ and the ring signature $\sigma$ as input and returns either \verb"TRUE" or \verb"FALSE".
\end{itemize} 
\end{definition}
\subsection{Properties} A ring signature must satisfy the usual correctness and unforgeability property. A fairly generated ring signature must be accepted as  valid with higher probability; and it must be infeasible for any other user to generate, except a very little probability $\epsilon$, a valid ring signature with the ring he does not belong to.
\paragraph{} Signature must be anonymous, so that no verifier should  be able to guess the actual signer's identity with probability greater than ${1 \over r} + \epsilon$, where $r$ is ring size and $\epsilon$ is however small.
\paragraph{} Since the possible signer set is randomly chosen and is not predetermined, it should be a part of signature.
\subsection{Combining Function} The concept of \textit{Ring signature} is derived from an abstract concept called \textit{Combining Function}.
\begin{definition}
A combining function $C_{k, v}(y_1, y_2, ..., y_n)$ takes as input a key $k$, an initialization value (also refereed as glue value) $v$, and an arbitrary values $y_1, y_2, ..., y_n~ \in ~ \{0, 1\}^b$. It produces $z \in \{0, 1\}^b$, such that for any fixed values $k, v$, any index $s$ and fixed value of $\{y_i\}_{i \neq s}, C_{k, v}$ is a permutation over $\{0, 1\}^b$, when seen as a function of $y_s$. This permutation is efficiently computable as well as its inverse.
\end{definition}
\subsection{Rivest et al.'s Ring signature scheme}
\subsubsection{Ring signature generation:}
Given a message $m$ to be signed, signer's secret key $S_k$, and the possible signers' public keys sequence $P_1, P_2, ..., P_r$ of all ring members, the signer computes the ring signature as follows.
\begin{enumerate}
\item \textbf{Choose a key} : The signer computes, using a publicly known hash function $h$ \[k = h (m, P_1, P_2, ..., P_r)\]
\item \textbf{Pick a random glue value} : The signer picks an initialization value $v {\in}_R \{0, 1\}^b$
\item \textbf{Pick random $x_i$'s} : The signer picks an random $x_i$ for all other ring members uniformly and independently from $\{0, 1\}^b$, and computes $y_i = g_i(x_i)$ .
\item \textbf{Formation of ring} : The signer solves the ring equation for $y_k$:
\[C_{k, v}(y_1, y_2, ..., y_n) = v \]
and using knowledge of his trapdoor he gets $x_k$ from $y_k$.
\item \textbf{Output the ring signature} : The ring signature on message $m$ is defined to be the $(2r + 1)$ tuple: \[(P_1, P_2, ..., P_r; v; x_1, x_2, ..., x_r)\]
\end{enumerate}

\subsubsection{Ring signature verification:}
On receiving $(P_1, P_2, ..., P_r; v; x_1, x_2, ..., x_r)$ as the ring signature  on message $m$, the verifier can verify as follows.
\begin{enumerate}
\item \textbf{Apply trapdoor permutation} : verifier computes for each $i$, $y_i = g_i(x_i)$
\item \textbf{Key computation} : computes $k = h (m, P_1, P_2, ..., P_r)$
\item \textbf{Verify the ring equation} : The verifier checks if
\[C_{k, v}(y_1, y_2, ..., y_n) = v\]
If the ring equation is satisfied, the verifier accepts the signature as valid otherwise rejects.
\end{enumerate}
\section{Bilinear Pairings}
Let ${\textbf{G}}_{1}$ cyclic additive group generated by $P$, whose order is a prime
q, and G2 be a cyclic multiplicative group of the same order q: A bilinear pairing is
a map  $e :{\textbf{G}}_{1} \times {\textbf{G}}_{1}\longrightarrow {\textbf{G}}_{2}$
with the following properties: \\
P1: \textit{Bilinear}: $e(aP, bQ) = {e(P,Q)}^{ab}$; \\
P2: \textit{Non-degenerate}: There exists $P,Q \in {\textbf{G}}_{1}$ such that $e(P,Q)
\neq 1$;\\
P3: \textit{Computable}: There is an efficient algorithm to compute $e(P,Q)$
for all $P, Q \in {\textbf{G}}_{1}$.
\paragraph{}
When the DDHP (Decision Diffie-Hellman Problem) is easy but the CDHP (Computational
Diffie-Hellman Problem) is hard on the group $G$; we call $G$ a Gap Diffie-Hellman
(GDH) group. Such groups can be found on supersingular elliptic curves or
hyperelliptic curves over a finite field, and the bilinear parings can be derived from
the Weil or Tate pairing. \cite{Bon01,ChC03,Hes02}

\section{Proposed ID-based Ring Signature Scheme (IDBRS)}
\subsubsection{\textbf{Setup :}} Let $P$ is a generator of ${\textbf{G}}_{1}$; $ e : {\textbf{G}}_{1} \times {\textbf{G}}_{1}\longrightarrow {\textbf{G}}_{2}$ is a bilinear pairing. ${H}_{1} : {\{0, 1\}}^{*} \longrightarrow
{z}_{q}^*$, ${H}_{2} : {\{0, 1\}}^{*} \longrightarrow {\textbf{G}}_{1}$ and ${H}_{3} : {\textbf{G}}_{2} \longrightarrow {z}_{q}^* $ are cryptographic hash functions. Key Generation Center (KGC) chooses a random number $s \in Z_q^*$ and sets $P_{Pub} = sP$. The KGC publishes the system parameters $\{G_1, G_2, e, q, P, P_{Pub}, H_1, H_2\}$ and keeps s as the master key.
\subsubsection{\textbf{Extract :}} An user submits its identity information $ID_k$ to KGC. KGC publishes the public key $Q_{k} = H_2(ID_k)$ and returns $S_k = s Q_k$ to the user as his/her private key.
\subsubsection{\textbf{Ring signature generation:}}
Given a message $m$ to be signed, signer's secret key $S_k$, and the possible signers' public keys sequence $L = (ID_1, ID_2, ..., ID_r)$ of all ring members, the signer computes the ring signature as follows.
\begin{enumerate}
\item \textbf{Choose a key} : $K = H_1 (m || L).$
\item \textbf{Pick a random glue value} : The signer picks a random $A \in G_1$ and computes the initialization value \[v = c_k = e(A, P)^K\]
\item \textbf{Pick random $T_i$'s} : The signer picks a random $T_i$ for all other ring members uniformly and independently from $G_1$, and computes \[c_{i+1} = [e(P_{Pub}, H_3(c_i) Q_i).e(T_i, P)]^K\] .
\item \textbf{Formation of ring} : The signer solves the ring equation for $y_k$. when $i = k$, we get
\[c_{k+1} = [e(P_{Pub}, H_3(c_k) Q_k).e(T_k, P)]^K = v\]
On solving this ring equation we get
\[T_k = A - H_3(c_k) S_k\]
Now compute $T = \Sigma~T_i$
\item \textbf{Output the ring signature} : The ring signature on message $m$ is the tuple \[(L; c_1, c_2, ..., c_r; T)\]
\end{enumerate}
\subsubsection{Ring signature verification:}
On receiving the ring signature $(L; c_1, c_2, ..., c_r; T)$ on message $m$, the verifier can verify as follows.
The verifier computes
\[K = H_1 (m || L).\]
and checks if
\[ {\Pi}_i c_i = [e(P_{Pub}, {\Sigma}_i (H_3(c_i)Q_i)).e(T, P)]^K\]
If the equation is satisfied, the verifier accepts the signature as valid otherwise rejects.
\section{Analysis of IDBRS}
\subsection{Correctness}
From ring signature generation protocol --
\[c_{i+1} = [e(P_{Pub}, H_3(c_i) Q_i).e(T_i, P)]^K\]
\[{\Pi}_{i=0}^r c_{i+1}= {\Pi}_{i=0}^r [e(P_{Pub}, H_3(c_i) Q_i).e(T_i, P)]^K\]
\[{\Pi}_{i=0}^r c_{i+1}= [{\Pi}_{i=0}^r e(P_{Pub}, H_3(c_i) Q_i). {\Pi}_{i=0}^r e(T_i, P)]^K\]
\[{\Pi}_{i=0}^r c_{i+1}= [ e(P_{Pub}, {\Sigma}_{i=0}^r (H_3(c_i) Q_i)). e({\Sigma}_{i=0}^r T_i, P)]^K\]
\[ {\Pi}_i c_i = [e(P_{Pub}, {\Sigma}_i (H_3(c_i)Q_i)).e(T, P)]^K\]
which hold true, since we have $c_{r+1}=c_0$
\subsection{Security}
The proposed ID-based ring scheme holds unconditionally signer-ambiguity, as all $T_i$
but $T_k$ are taken randomly from $G_1$: In fact, at the starting point, the $T_k$ is also
distributed uniformly over $G_1$, since $A$ is randomly chosen from $G_1$. We fix a set of identities, denoted by $L$.

Suppose that $\mathcal{A}$ is an adversary whose identity $ID_{A}$ is not listed in $L$, but he wants to forge a valid ring signature. $\mathcal{A}$ can either forge a valid signature of a user whose identity $ID_k$ is listed in $L$ or executes the following experiment:
\begin{enumerate}
\item  $\mathcal{A}$ queries \verb"Extract" $q_E~,~(q_E > 0)$ times with known parameters and $ID_i$, which does not not belongs to $L$,for $i = 1, 2, ..., q_E$. The query \verb"Extract" returns the $q_E$ corresponding secret key such that  $S_i = s H_2(ID_i) = s Q_i$ .
\item He Chooses randomly an integer $c_0 \in Z_q^*$
\item He runs ring signature generation protocol's third step for $i = 0, 1, ..., r-2$,  where $r =|L|$
\item Assigns $c_{0} = [e(P_{Pub}, H_3(c_{k-1}) Q_{k-1}).e(T_{k-1}, P)]^K $ 
\item Outputs the ring signature $(L; c_1, c_2, ..., c_r; T)$
\end{enumerate}
After running Step 1 of the above experiment, $\mathcal{A}$ gets $\{S_1, S_2, ..., S_{q_E}\}$, a set of secret keys . Suppose he gets a pair $(ID_m, S_m)$ such that $H_2(ID_m) = H_2(ID_j)$, where $ID_J \in L$, then he can forge a valid ring signature. But since $H_2$ is random oracle and \verb"Extract" generates random numbers with uniform distributions. This implies that $\mathcal{A}$ gets nothing from query results. $H_3$ is random oracle and all Ti are taken randomly from $G_1$. This implies that the probability of $c_{0} = [e(P_{Pub}, H_3(c_{k-1}) Q_{k-1}).e(T_{k-1}, P)]^K $ to be true is $1 \over q$. So we can say that the proposed scheme is non-forgeable. 
\subsection{Efficiency}
The proposed ring signature scheme works under the environment of supersingular elliptic curves or hyperelliptic curves. The essential operation in our ID-based signature schemes is to compute a bilinear pairing. 
\paragraph{} We denote by $P$, the cost of computation a bilinear pairing, $A_{G_1}$ the cost of addition in $G_1$, $M_{G_1}$ cost of multiplication in $G_1$, $M_{G_2}$ cost of multiplication in $G_2$ and cost of multiplication in $Z_q$ by $M_{Z_q}$. The cost of hashing is denoted by $H$. We shall not consider exponentiation as it can be reduced in addition  in $G_1$. We ignore the cost of computation of $H_2(ID)$

\begin{table}
  \centering
  \caption{Comparison of computational cost with existing schemes.}\label{Tab1}
  \begin{tabular}{|c|c|c|}
    \hline
    
       & Signature Generation &  Verification \\ 
     \hline
     Zhang's Scheme & $(2n -1)P + nH + nA_{G_1}  $ & $2nP + nH + nM_{G_1} + nM_{G_2}$ \\
                    & $+ nM_{G_1}+ (n-1)M_{G_2}$ &  \\
      \hline
     Lin's Scheme & $(2n -1)P + H + nA_{G_1}  $ & $2P + H + (n-1)A_{G_1}$ \\
                    & $+ (2n-1)M_{G_1}+ nM_{G_2}$ & $+ (n+1)M_{G_1} + n M_{G_2}$ \\
      \hline
    Proposed Scheme & $(2n -1)P + nH + (n+1)A_{G_1}  $  & $2P + (n+1)H + (n+1)A_{G_1}  $ \\
        & $+ 2nM_{G_1}+ (n-1)M_{G_2}$ & $+(n+1)M_{G_1} + (n-1)M_{Z_q} +M_{G_2} $ \\
     \hline
  \end{tabular}
\end{table}
In our opinion both the first two schemes discussed in \textbf{Table \ref{Tab1}} are having inconsistency in the computational procedure. As in Zhang's scheme \cite{ZK02}, in initialization, $c_{k+1} = H(L||m||e(A, P))$ has been computed, which is incorrect. $H$ is defined in their paper as $H:\{0, 1\}^* \rightarrow Z_q$. But in computation of $c_{k+1}$, the pairing $e(A, P)$ had used, which belongs to $V$ (according to their notation), not to $Z_q$. This shows their $c_{k+1}$ computation is taken incorrectly. We may remove this inconsistency by applying a newly defined hash function as $H_4:G_2 \rightarrow \{0, 1\}^*$. If we modify their scheme's in this way, computational cost of their scheme increases by a factor $nH$ in signature phase and by a factor of $nH$ in verification phase. A very similar mistake is in Lin's scheme \cite{Lin-Wu:2003:117} is made. In Equation 3, they have computed $c_{a+1} = e(A, P)$. This implies that the  $c_{a+1}$ is an element of $G_2$, but they have treated it as element of $Z_q$ in equation 4 and also in equation 5. (If $P \in G_2$ and $Q \in G_1$ then $P.Q$ is not defined.). If we define a hash function $H_5:G_2 \rightarrow Z_q$, computational cost of this scheme is also increased by a factor $nH$ in signature generation phase and also $nH$ in verification phase. Our scheme does not contain such inconsistency and also is more efficient. 
\section{ID-based Proxy Ring Signature Scheme (IDBPRS)}
\subsection{Delegation Function due to Zhang et al. \cite{ZNL03}} Here an original signer with secret
key- public key pair $({x}_{o}, {PK}_{o})$ wants to delegate signing power to proxy signer with secret key- public key pair $({x}_{p}, {PK}_{p})$. System parameters
are $\{{G}_{1}, {G}_{2}, e, q, P, {H}_{1}, {H}_{2} \}$. The original signer runs the
following protocol -
\begin{itemize}
\item The original signer prepares a warrant message consist of explicit description
of the delegation relation. warrant message also contains some identity information of
the proxy signer. \item The original signer computes ${x}_{ow} = {x}_{o} {H}_{2}(w)$
and sends $({x}_{ow}, w)$ to the proxy signer. \item Proxy signer checks $e({x}_{ow},
P) = e({H}_{2}(w), {PK}_{o})$. If it holds, he computes then ${x}_{w} = {x}_{ow} +
{x}_{p}{H}_{r}(w)$.
\end{itemize}
Above protocol can be regarded as \textbf{PKGen} (Proxy key generation protocol) in
proxy signature scheme. In this delegation proxy signer will use ${x}_{w}$ as secret
key and ${PK}_{o}+{PK}_{p} $ as public key. Now proxy signer may use any ID-based
signcryption scheme from pairing (takes the ID public key as ${H}_{2}(w)$ and secret
key ${x}_{w}$ and the public key of trusted authority as ${PK}_{o}+{PK}_{p})$ to get
proxy signcryption scheme. Security of above protocol is discussed in \cite{ZNL03}.
\section{A New Proxy Ring Signature Scheme from Pairings}
\textbf{[Setup]} The system parameters \verb"params" = \{${\textbf{G}}_{1},
{\textbf{G}}_{2},e ,q,P,{H}_{1}, {H}_{2}\}$ Let Alice be the original signer with
public key ${PK}_{o} = {s}_{o} P$ and private key ${s}_{o}$, and $L = \{{PS}_{i}\}$ be
the set of proxy signers with public key ${PK}_{{p}_{i}} = {s}_{{p}_{i}} P$ and
private key ${s}_{{p}_{i}}$.\\\textbf{[Proxy Key Generation]} The original signer
prepares a warrant $w$, which is explicit description of the delegation relation. Then
he sends $(w, {s}_{o} {H}_{2}(w))$ to the proxy group L. Each proxy signer uses his
secret key ${S}_{{p}_{i}}$ to sign the warrant $w$ and gets his proxy key ${S}_{i} =
{s}_{o} {H}_{2}(w) + {s}_{{p}_{i}} {H}_{2}(w)$.\\\textbf{[Proxy Ring Signing]}For
signing any message $m$, the proxy signer ${PS}_{i}$ chooses a subset $L^{\prime}
\subseteq L$. Proxy signers's public key is listed in $L^{\prime}$. Now to sign he/
she perform following operations:

\begin{itemize}
\item{\textbf{Initialization}}: Choose randomly an element $A \in {\textbf{G}}_{1}$, compute
\begin{equation}
{c}_{k+1} = e(A, P)
\end{equation}
%
\item{\textbf{Generate forward ring sequence}} For $i = k + 1,k + 2,....k+(n -1)$ choose
randomly ${T}_{i} \in {\textbf{G}}_{1}$ and compute
\begin{equation}
{c}_{i+1} = e{({PK}_{o}+{PK}_{{p}_{i}}, H_3({c}_{i}){H}_{2}(w))}^{{H}_{2}(m \parallel L)}.
e(Ti, P)
\end{equation}
\item{\textbf{Forming the ring}}: Let ${R}_{n}={R}_{o}$. Then, ${PS}_{i}$ computes
\begin{equation}
{T}_{i} = A - {h}_{2}(m \parallel L) H_3({c}_{i}){S}_{i},
\end{equation}
\begin{equation}
T = {\Sigma}_{i=1}^{n} {T}_{i}
\end{equation}
\item{\textbf{Output}}: Finally, Let ${c}_{n} = {c}_{0}$. The resulting ring signature for a
message $m$ and with ring member specified by $L^{\prime}$ is the
$(n+1)$-tuple:$({c}_{1}, {c}_{2}, ..., {c}_{n}, T)$
\end{itemize}
\paragraph{}
\textbf{[Verification]} Given message $m$, its ring signature $({c}_{1}, {c}_{2}, ...,
{c}_{n}, T)$, and the set $L^{\prime}$ of the identities of all ring members, the
verifier can check the validity of the signature by the testing if:

\begin{equation}
{\Pi}_{i=1}^{n} {c}_{i} = e{({PK}_{o}+{PK}_{{p}_{i}}, {\Sigma}_{i=1}^{n} H_3({c}_{i})
{H}_{1}(w))}^{{H}_{2}(m \parallel L)}.e(T,P)
\end{equation}

\section{Analysis}

\textbf{Key Secrecy} In computing user ${P}_{i}$'s private key ${S}_{i}$ from the
corresponding public key ${PK}_{o} +{PK}_{{P}_{i}}$ requires the knowledge of original
signer's private key ${s}_{o}$ and proxy signer's private key ${s}_{{p}_{i}}$.
According to definition these keys are protected under the intractability of DLP in
${G}_{1}$ as ${PK}_{o}= {s}_{o}P$ and ${PK}_{{P}_{i}} = {s}_{{p}_{i}}P$.\\
\textbf{Signer ambiguity} In a valid proxy ring signature $({c}_{1}, {c}_{2}, ...,
{c}_{n}, T)$ with proxy group $L^{\prime}$ generated by ${PS}_{i}$ all ${c}_{i}$'s
are computed by eq 2. Since ${T}_{i} \in {G}_{1}$ is chosen uniformly at random, each
${c}_{i}$ is uniformly distributed over ${G}_{2}$. Thus, regardless who the actual
signer is and how many ring members involved $({c}_{1}, {c}_{2}, ..., {c}_{n})$ biases
to no specific ring member.
Other discussion are very similar as in previous sections.
\section{Conclusion}
In this paper we proposed a new ID-based ring signature scheme from bilinear pairings. This scheme removes deficiencies in existing schemes. In this paper we proposed a new and a proxy ring signature scheme which, whenever proxy signer want to sign message on behalf of the original signer
provide anonymity. The proposed scheme is more efficient than the scheme of Zhang
et al.'s, especially for the pairing operation required in the signature verification.
This proxy ring signature scheme is more efficient for those verifiers who have
limited computing power.
\bibliographystyle{amsplain}

\end{document}